\documentclass{emulateapj}
\usepackage{times,common}
\usepackage{pdfpages}

\begin{document}
\defcitealias{owers09}{ONC}

\markboth{Cheng-Jiun Ma et al.}{A1201 merger}
\title{Abell 1201: a Minor merger at second core passage}

\author{
Cheng-Jiun Ma\altaffilmark{1,2},
Matt Owers\altaffilmark{3,4},
Paul E. J. Nulsen\altaffilmark{1},
Brian R. McNamara\altaffilmark{1,2,5},
Stephen S. Murray\altaffilmark{1,6},
Warrick J. Couch\altaffilmark{4}
}
\altaffiltext{1}{Harvard-Smithsonian Center for Astrophysics, 60
  Garden St., Cambridge, Massachusetts, 02138-1516, USA} 
\altaffiltext{2}{Department of Physics \& Astronomy, University of
  Waterloo, 200 University Ave. W., Waterloo, Ontario, N2L 3G1,
  Canada} 
\altaffiltext{3}{Australian Astronomical Observatory, PO Box 296,
  Epping, NSW 1710, Australia}
\altaffiltext{4}{Centre for Astrophysics \& Supercomputing, Swinburne
  University of Technology, PO Box 218, Hawthorn, VIC 3122, Australia}
\altaffiltext{5}{Perimeter Institute for Theoretical Physics, 31
  Caroline St. N., Waterloo, Ontario, N2L 2Y5, Canada} 
\altaffiltext{6}{Department of Physics and Astronomy, Johns Hopkins
  University, 3400 North Charles St., Baltimore, MD 21205, USA}

\begin{abstract}

We present an analysis of the structures and dynamics of the merging cluster Abell~1201, which has  two sloshing cold fronts around a cooling core, and an offset gas core approximately 500\,kpc northwest of the center. New \chandra\ and \xmm{} data reveal a region of enhanced brightness east of the offset core, with breaks in surface brightness along its boundary to the north and east. 
This is interpreted 
as a tail of gas  stripped from the offset core. Gas in the offset core and the tail is distinguished from 
other gas at the same distance from the cluster center chiefly by having higher density, hence lower entropy. 
In addition, the offset core shows marginally lower temperature and metallicity than the 
surrounding area. The metallicity in the cool core is high and there is an abrupt drop in metallicity across the southern cold front. 
We interpret the observed properties of the system, including the placement of the
cold fronts, the offset core and its tail in terms of a simple merger scenario. The offset core is the remnant of a merging 
subcluster, which first passed pericenter southeast of the center of the primary cluster 
and is now close to its second pericenter passage, moving at $\simeq 1000\rm\,km\,s^{-1}$.  
Sloshing excited by the merger gave rise to the two cold fronts and 
the disposition of the cold fronts reveals that we view the merger from close to the plane of 
the orbit of the offset core. 

\end{abstract}


\section{Introduction}\label{sec:intro}

In hierarchical structure formation models, galaxy clusters are formed
by mergers of smaller systems, including other groups and clusters,
taking place over the age of the universe \citep[e.g.][]{springel06}.
Mergers between galaxy clusters are among the most energetic events in
the universe \citep{sarazin02}.  Although the merger rate is now
decreasing with time, evidence of recent and ongoing mergers is still
commonly found in clusters.  Cluster mergers can induce pronounced
observable features, particularly in the X-rays.  High resolution X-ray
observations with \chandra{} and \xmm{} provide a unique means to
study cluster mergers.  Combined with galaxy redshifts from optical
spectroscopy and mass distributions from lensing, they provide a
valuable tool for studying merger dynamics and testing our
understanding of cluster physics.

A notable discovery from X-ray astrophysics in the past decade is the
phenomenon of cold fronts in clusters \citep[][]{markevitch00,
  vikhlinin01a}.  These are contact discontinuities where there is an
abrupt change in the entropy of the gas \citep{markevitch07}.
Simulations \citep[e.g.][]{churazov03, tittley05, ascasibar06,
  poole06, poole08, zuhone10, roediger11} show that cold fronts can be
induced in more than one way during cluster mergers.  Cold fronts of
the ``remnant core type'' occur at the interface between the core of
an infalling subcluster and the warmer ICM of the primary cluster
\citep[e.g.~1E0657-56;][]{markevitch02}.  Another type, called
``sloshing'' cold fronts, result from gas motions in a cluster core
induced by the gravitational perturbation of an infalling subcluster
\citep[e.g.~Abell 1795;][]{markevitch01}.  Simulations
\citep[e.g][]{ascasibar06, zuhone10} show that the ``sloshing" type of
cold fronts appear $\sim 0.3$\,Gyr after pericentric passage of a
subcluster.  This cold front moves outward and a second cold front can
appear on the opposite side of the center $0.6$\,Gyr or longer after
pericentric passage.  While the cold fronts continue to move outward,
additional cold fronts may appear on alternating sides of the center.
Viewed from many directions, the fronts are connected in a spiral
pattern.  Although the time scales vary from cluster to cluster, the
well-defined structure of alternating fronts and/or a spiral, are a
signature of sloshing induced by mergers.  These features have been
shown to be useful for constraining the dynamics and history of
mergers \citep[e.g.][]{johnson10,johnson11}.
 
Abell 1201 (A1201) is a typical example of a cluster with
sloshing cold fronts \citep[][\citetalias{owers09} hereafter; Owers et~al. 2011b]{owers09}.  It has a redshift of 0.168
\citep{struble99} and an X-ray luminosity of $L_{\rm
  X}(0.1-2.4\rm\,keV)=2.4\times10^{44}$\,erg\,s$^{-1}$
\citep{bohringer00, ebeling98}.  \citetalias{owers09} measured the redshifts of 321 member galaxies and found a
mean redshift of $z = 0.1673\pm0.0002$ and a velocity dispersion of
$778\pm36$\,km\,s$^{-1}$.  Using 21.5 ksec of ACIS-S data,
\citetalias{owers09} measured a global temperature of $5.3\pm0.3$\,keV
and an abundance of $0.34\pm0.10$ for A1201.  They also studied the
spatial and redshift distributions of the galaxies in A1201 and they
identified a remnant, offset core, located at the northwest end of a
bright X-ray ridge that runs through the cluster center \citep[see also][]{owers09b}.  \citet{edge03} analyzed strong lensing in A1201,
using the image of an arc located over the BCG, and found that the
ellipticity of the mass distribution in the cluster core exceeds that
of the optical isophotes of the BCG.

In this paper, we extend the study of \citetalias{owers09}, using deeper
\chandra{} and \xmm{} data to discuss the substructure and dynamics
of merging in A1201. The X-ray data reduction is discussed in
Section~\ref{sec:data}.  The substructures of A1201, including its cold fronts,
offset core, and tail are discussed in Section~\ref{sec:sb}.  These results
are interpreted in terms of a merger scenario presented in
Section~\ref{sec:discussion}. Section~\ref{sec:summary} is a short summary. 

For this analysis, we adopt a $\Lambda$CDM cosmology with
$h_0=0.7$, $\Omega_{\Lambda}=0.7$, and $\Omega_m = 0.3$.  In this
cosmology, $1\arcsec$ corresponds to 2.88\,kpc at a redshift of
$z=0.1673$.  Positions angles (PA) are measured counterclockwise from
the west. 

\begin{figure*}[th]
\epsscale{1.0} \plotone{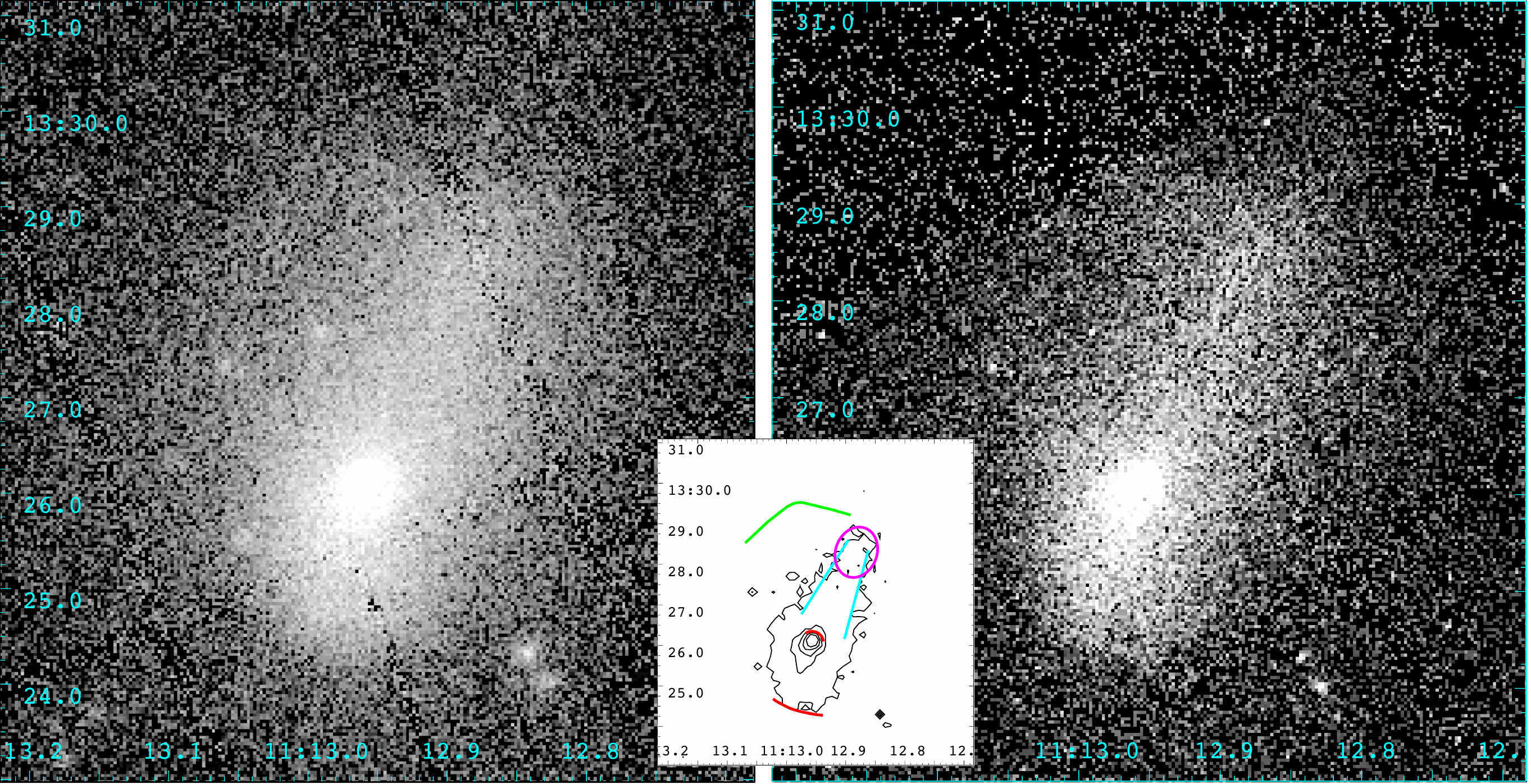} 
\figcaption{Exposure
  corrected, 0.5 -- 7 keV images from the EPIC data (Left) and from
  the ACIS data (Right).  Both images have been binned to
  $2\arcsec$ per pixel.  The substructures discussed in this paper
  are marked in color in the inset image, with X-ray contours in
  black: the two cold fronts (red), the offset core (magenta), the
  edge of the tail (green), and the ridge (cyan). 
} \label{fig:imagexmm}
\end{figure*}

\section{Data Reduction}\label{sec:data}

A1201 was observed using the European Photon Imaging Camera (EPIC) on
\xmm{} in December 2007 (ObsID 0500760101), and the Advanced CCD
Imaging Spectrometer (ACIS) on \chandra{} in April 2008 (ACIS-I; ObsID
9616) and in November 2003 (ACIS-S; ObsID 4216). The EPIC observations
were performed in full-frame mode using the medium filter, for a
total exposure time of 50\,ksec. The ACIS-I and ACIS-S observations were
performed in VFAINT mode for total exposure times of 47\,ksec and
40\,ksec, respectively.

\subsection{X-ray Imaging}\label{sec:data_image}

We reprocessed the EPIC data using SAS v10.0.0 with standard settings.
Light curves for the 10 -- 12 keV band for each of the three cameras
(PN, MOS1 and MOS2) were used to filter out periods when the count
rate exceeded the mean by $3\sigma$.  Images were produced for each
camera, using the method of \citet{carter07}.  Blank sky and
filter-wheel closed (FWC) datasets produced by the EPIC Blank Sky
team\footnote{http://xmm.vilspa.esa.es/external/xmm\_sw\_cal/background/blank\_sky.shtml.
  August 2010 version.} were used for this purpose.  Particle
background is first subtracted using FWC data, scaled by the counts in
regions outside each detector field of view.  The same procedure is
applied to the blank sky datasets, in principle, leaving only the
cosmic X-ray background (CXB).  Assuming that there are no source
photons in this region, counts in annuli 10 -- 12 arcmin from the
center of each detector are then used to scale the blank sky
backgrounds to the source exposures (for both, after subtraction of the
hard particle background) in order to remove CXB from the source data.
Allowing for some difference in the CXB spectrum between the blank sky
and source exposures, this procedure is carried out in the 4 energy
bands delimited by 0.5, 2.125, 3.750, 5.375, and 7.0\,keV.  The
background subtracted images are then divided by vignetted exposure
maps to obtain the image (left panel of Figure~\ref{fig:imagexmm}).

The ACIS-I (ObsID 9616) and ACIS-S (ObsID 4216) data sets were
reprocessed using the CIAO software package \citep[version
  4.3;][]{ciao}.  Following standard practice, light curves from
source free regions were used to filter out periods when the count
rate differed by more than $3\sigma$ from the mean.  The ACIS-S
observation is badly affected by a flare, which leaves only 21.5\,ksec
of useful exposure.  No significant flaring is seen in the ACIS-I
observation and the cleaned exposure time is 43\,ksec.  Background
subtraction was done using the standard blank sky datasets.  The
exposure map was generated for emission for an optically thin plasma
with a temperature of $kT = 5.3$\,keV, the global temperature of A1201
determined by \citetalias{owers09}.  The combined, 0.5 -- 7 keV,
background subtracted, exposure corrected image from the two data
sets, binned by a factor of 4, is displayed in the right panel of
Figure~\ref{fig:imagexmm}.

An inset image in Figure~\ref{fig:imagexmm} shows the regions discussed
in later sections of this paper.  The two cold fronts discussed by
\citetalias{owers09} and \citet{owers09b} are marked by red arcs.  The
offset core to the northwest of the cluster center is outlined by a
magenta ellipse.  The outer edge of the tail to the east of the offset
core is marked in green.  The ridge connecting the cool core
and the offset core is marked by two cyan lines.

\subsection{X-ray Spectral Analysis}\label{sec:data_spec}

All X-ray spectra were extracted in the energy range 0.5 to 7.0\,keV
and grouped to ensure a minimum of 20 counts per bin.  PN data were
not used in the spectral analysis, since the soft background cannot be
convincingly removed.

Spectra and response files for the ACIS-I and ACIS-S data were
assembled using the CIAO script ``specextract.''  Regions containing
X-ray point sources were identified using the CIAO ``cell\_detect''
task with a $3\sigma$ threshold and excluded from all extracted
spectra.  The ACIS-S observation was performed at the standard focal
plane temperature of $-120\degr$C, but the focal plane was slightly
warmer ($-118.7\degr$C) for the ACIS-I observation.  This is expected
to degrade the calibration accuracy.  However, our spectral fits
reveal no significant discrepancy and we assume that any calibration
errors are insignificant.  Background spectra were created using the
standard blank sky data described in \citet{markevitch98}.  For the
MOS data, spectra were extracted using the SAS ``evselect'' task.  The
point sources identified in the ACIS-I data were excluded, but using
larger, $9\arcsec$ radius, apertures.  Background spectra were created
using a double background subtraction approach similar to that used
to make the images.  Briefly, the background spectrum is assumed to be
the sum of two components, a hard particle component, which is
determined by scaling FWC data, and a cosmic background component.
The cosmic X-ray background is determined by scaling the residual
blank sky spectrum, after subtraction of the particle background, to
make the net spectrum in the source exposure zero in the annulus
between $10\arcmin$ and $11\arcmin$ centered on the cluster.

\begin{deluxetable}{lcccc}
\tabletypesize{\scriptsize}
\tablewidth{0pc}
\tablecolumns{5} 
\tablecaption{Global properties\label{table:global_temp}}
\tablehead{ 
\colhead{Data} & \colhead{kT (keV)} & \colhead{Z} & \colhead{$\chi^{2}$/DoF} & \colhead{counts\tablenotemark{a}}
}
\startdata
ACIS-I        & $5.56\pm 0.2$ & $0.31\pm 0.1$ & $205/266$  & $17762$   \\
ACIS-S      &  $5.29\pm 0.3$ & $0.30\pm 0.1$ & $194/225$ & $16499$    \\
ACIS          &  $5.43\pm 0.2$ & $0.32\pm 0.07$ & $404/493$ & \nodata   \\
MOS          & $5.16\pm 0.2$  & $0.28\pm 0.05$ & $532/537$ & $38353$ \\
ACIS+MOS& $5.31\pm 0.15$ & $0.29\pm 0.039$ & $853/958$ &\nodata \\
Owers+08 & $5.3\pm 0.3$   & $0.34\pm 0.10$ &                 &
\enddata
\tablenotetext{a}{Net 0.5 -- 7\,keV counts, after background subtraction.}
\end{deluxetable}

To determine a global temperature and metallicity, spectra were
extracted from the elliptical region defined in Figure 4 of
\citetalias{owers09}.  These were fitted with an absorbed (WABS),
optically thin, thermal plasma model \citep[MEKAL;][]{mewe86}, using
Sherpa.  The absorbing column density of neutral hydrogen was fixed at
$N_{\rm H}=1.61\times10^{20}$\,cm$^{-2}$, allowing for foreground gas
\citep{dickey90}.  Spectra from the different instruments were fitted
jointly using the same model parameters. The normalizations
were not tied together, since their values are sensitive to
assumptions about the spatial distribution of the X-ray emission and
relative calibration of the detectors.  Table~\ref{table:global_temp}
lists best fitting parameters for a few instrument combinations,
together with the results of \citetalias{owers09} for comparison.
Uncertainties are 90\% confidence limits.  The net photon counts for
the four detectors, given in the last column of the table, are
similar. Thus, the total net photon count in the spectra used here is
roughly four times greater than used by \citetalias{owers09}.
\citet{nevalainen10} found that temperatures determined from EPIC (PN
and MOS) data are typically 10\% lower than temperatures for the same
region determined from ACIS data when fitting in a broad band (0.75 --
7\,keV).  This is roughly consistent with the temperature differences
seen in our data (Table~\ref{table:global_temp}).

\begin{figure*}[h]
\epsscale{0.9}
\plottwo{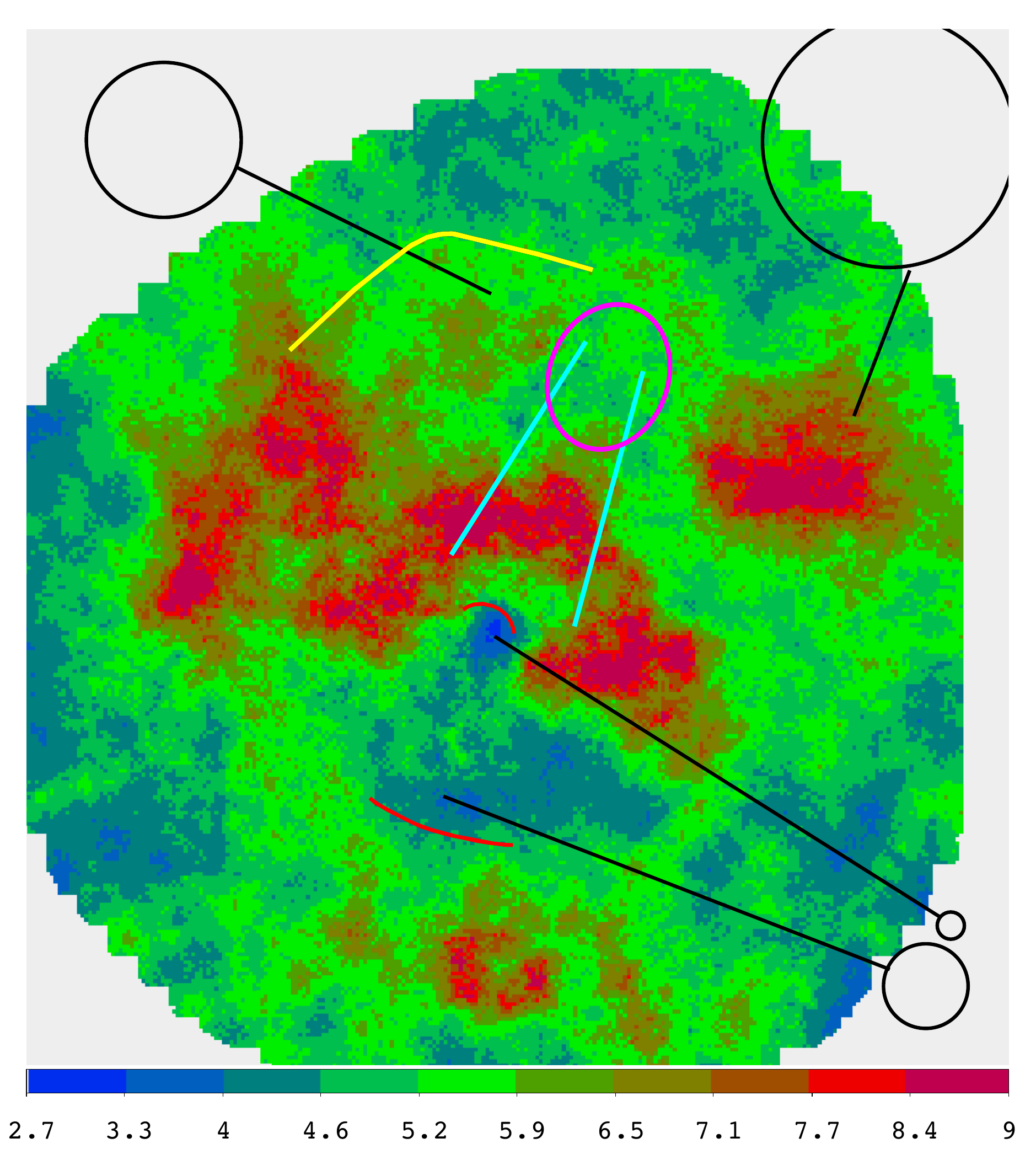}{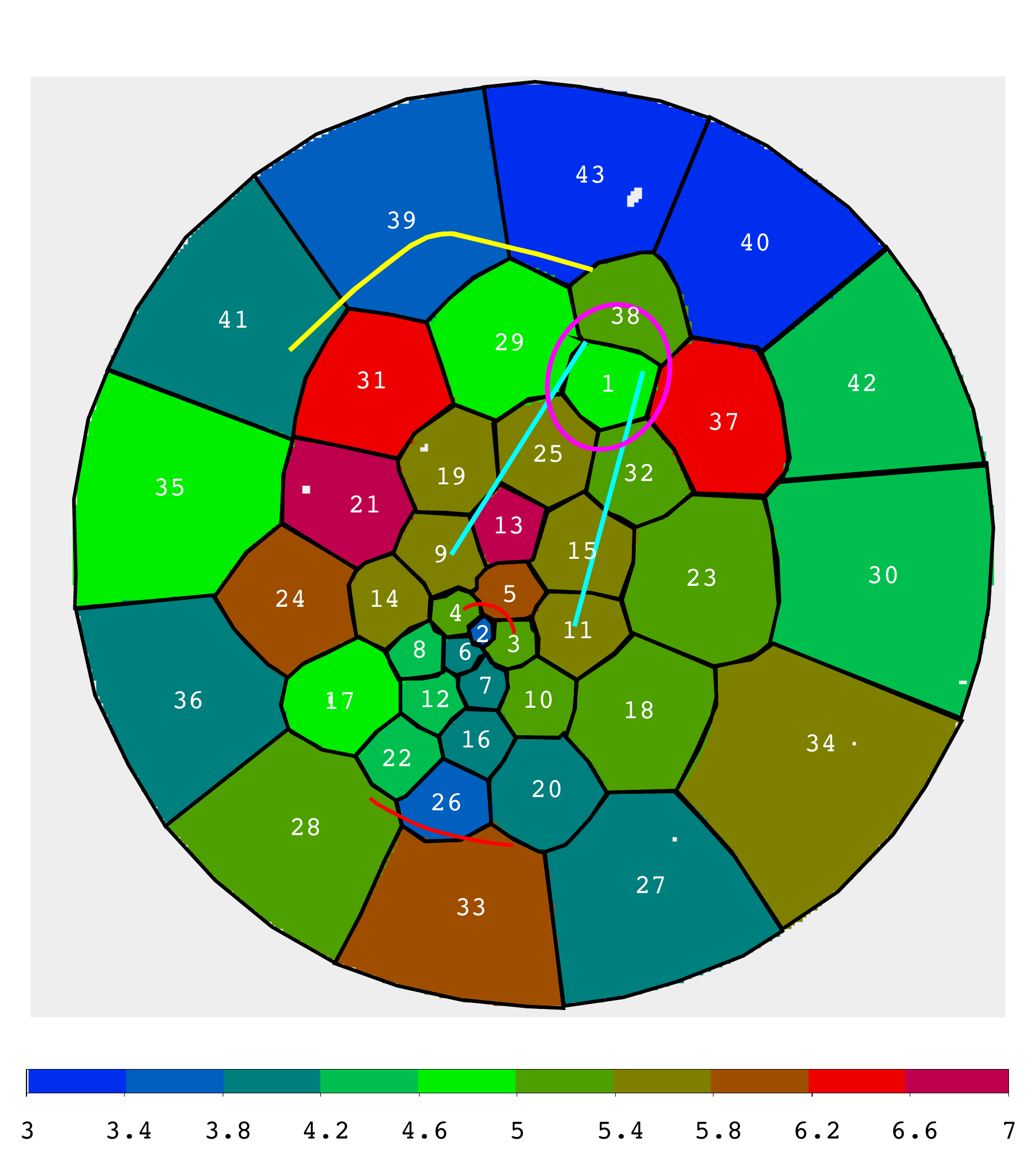}
\figcaption{{\em Left:} High resolution temperature map. The temperature for each pixel is determined from the spectrum extracted within a circle containing 1000 background subtracted counts. The details are described in \S\ref{sec:data_tempmap}. The median uncertainty (68\% confidence) ranges from 0.75\,keV at $kT \sim 4$\,keV to 1.6\,keV at $T \sim 7$\,keV. The black circles mark the extraction regions at some sample pixels. {\em Right:} WVT temperature map from
  the combined ACIS+MOS data.  Spatial bins were defined by applying
  the WVT binning algorithm to the ACIS-I data, with the signal to
  noise ratio set to 25.  Temperature uncertainties are about 0.5\,keV
  at $kT \sim 4$\,keV and 0.7\,keV at $kT \sim
  6$\,keV.    Substructures of interest are marked on both panels, as in
  Figure~\ref{fig:imagexmm}. \label{fig:matt_hires_tmap}} 
\end{figure*}

\begin{figure*}[t]
\epsscale{0.9}
\plottwo{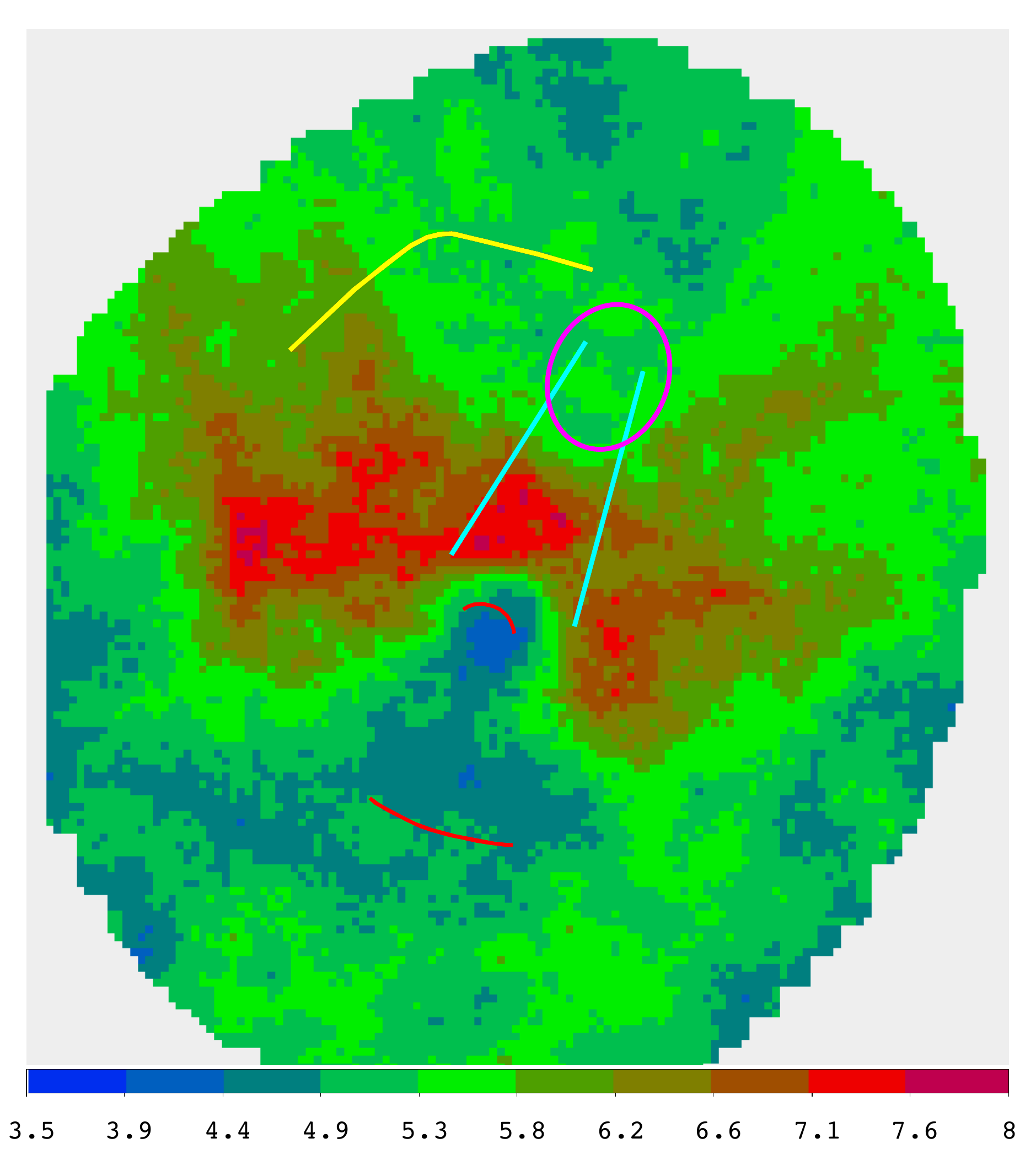}{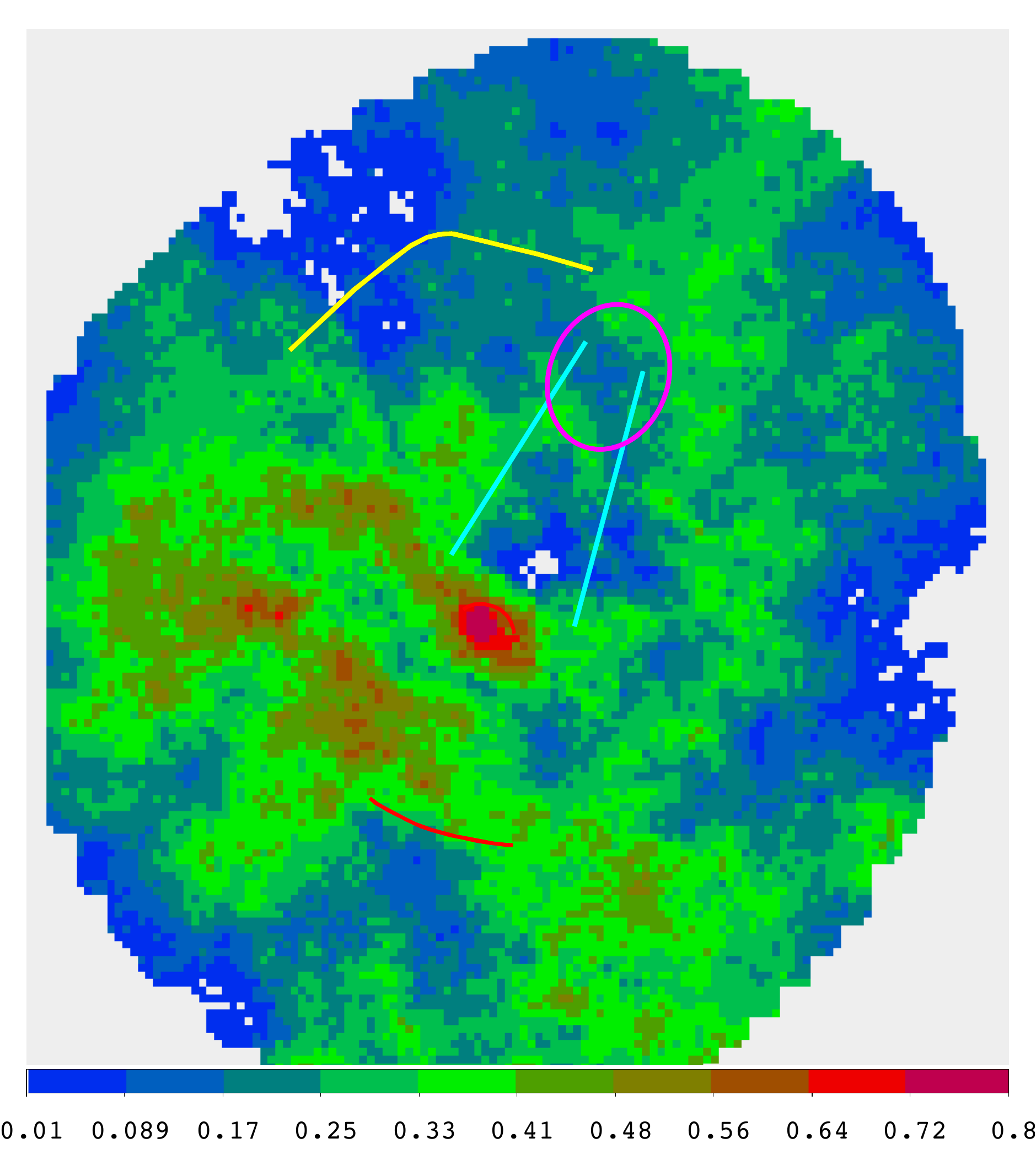}
\figcaption{ACIS temperature ({\em Left}) and abundance ({\em Right}) maps. The temperature maps differs from the left panel of Figure~\ref{fig:matt_hires_tmap} only by using spectra from regions containing 3000 background subtracted counts around each pixel. The same spectra were used to determine the abundances plotted in the right panel. The  median temperature uncertainty (68\%
  confidence) ranges from about 10\% for $\rm kT<6$\,keV to 20\% for $\rm kT>6$\,keV. In the bright regions, the uncertainties on the abundance measurements are $\sim 0.1$ and rise to $\sim 0.2$ in the fainter regions.  Substructures of interest are marked, as in Figure~\ref{fig:imagexmm} and Figure~\ref{fig:matt_hires_tmap}. \label{fig:matt_tmap}} 
\end{figure*}

\subsection{Temperature and Abundance Maps}\label{sec:data_tempmap}

\citetalias{owers09} employed the broad band method of
\citet{markevitch00} to make a temperature map from the ACIS-S data. 
An improved map, made by applying the methods described in \citet{randall08} and \citet{owers11a} to the combined ACIS-I and ACIS-S data, is shown in the left panel of Figure~\ref{fig:matt_hires_tmap}. Briefly, the $0.5-7.0$\,keV, background subtracted images for ACIS-I and ACIS-S data were coadded and binned by a factor of 4. At each pixel in the binned image, a spectrum was extracted from a circular region with radius set such that there are 1000 background-subtracted counts. The radius of the circles ranges from $6.7\arcsec$ at the center of cluster to $\sim 100\arcsec$ at the boundary of the map. The size of some extraction regions were marked with black circles in the left panel of Figure~\ref{fig:matt_hires_tmap}.  Weighted response files were extracted from a more coarsely binned image with binning factor of 32. The spectra were grouped to obtain at least 20 counts per channel. The data were fitted with an absorbed MEKAL model with the abundance fixed at $\rm Z=0.32$, i.e. the global value for the combined ACIS-S and ACIS-I fit in Table~\ref{table:global_temp}.

A second temperature map was generated by fitting spectra binned into regions defined by a weighted Voronoi tesselation \citep[WVT;][]{wvt} of the ACIS-I data.  The target signal to noise ratio for the
WVT binning was set to 25, so that each spatial bin contains about 625 counts.  
Using the same regions, the spectra from all four data sets (ACIS-I, ACIS-S, MOS1, and MOS2) were extracted. Since the net 0.5 --
7\,keV count is comparable for all four data sets.  Thus, the total net count for each spectral region
lies in the range $1800$ to $2800$, sufficient for robust temperature
measurements.  The signal to noise ratio for the WVT algorithm was
chosen so that the smallest regions have a size comparable to the
spatial resolution of XMM-Newton. Spectra were fitted using the absorbed MEKAL model as described above. Results of fitting spectra for the
ACIS and MOS data separately and jointly were compared.  Discrepancies
between ACIS and MOS temperatures comparable to those discussed in
Section~\ref{sec:data_spec} are seen, but the values generally agree within
1$\sigma$ statistical uncertainties.  The temperature map in the right
panel of Figure~\ref{fig:matt_hires_tmap} shows results for the joint fit.

We present both temperature maps in Figure~\ref{fig:matt_hires_tmap}, since
they have complementary properties.  The maps in the left 
panel retains more spatial information, since it does not require the
data to be binned on an artificial grid.  However, the results in adjacent
pixels are correlated over a distance that is not readily discerned
from the maps.  By contrast, the WVT map gives independent temperatures and abundances
for well-defined regions, at the expense of binning together regions
that may have disparate properties.  The two temperature maps are
broadly in agreement with one another.  The primary cluster core
(regions 2 of the bottom panels) is cool, with the cool gas
extending southward.  A stream of hotter gas crosses the bright ridge in the center of both maps.  The
subcluster core at the northern end of the ridge (region 1) is also marginally cooler than the surrounding regions.

The method used to make the high resolution temperature map of Figure~\ref{fig:matt_hires_tmap} was used with coarser binning to obtain the temperature (left) and abundance (right) maps of Figure~\ref{fig:matt_tmap}. In detail, the ACIS-I and ACIS-S data were binned by a factor of 8 and the spectrum at each pixel was extracted from a circular region containing at least 3000 background subtracted counts. The spectra were grouped to obtain at least one counts per channel. The minimal grouping is adopted to avoid diluting the 6.7\,keV iron line. Temperatures and abundances were fitted to an absorbed MEKAL model using the Cash statistics. 

The abundance of the cool core is high. The inner and outer cold fronts are revealed in both the temperature and abundance maps, with the abundance on the cooler side being higher than on the hotter side of both fronts. The ridge (between the cyan lines of Figure~\ref{fig:matt_tmap}), as well as the offset core (the magenta ellipse), shows a marginally low abundance.

\subsection{Deprojected Gas Properties}\label{sec:data_densitypfl}

An ``onion peeling'' deprojection method was used to obtain radial
temperature and density profiles for the gas of A1201
\citep{david01,nulsen05}.  Assuming that the gas distribution is
spherical and that the gas densities and temperatures are constant in
spherical shells, the algorithm uses spectral parameters fitted to the
outer layers to calculate the spectra of outer shells projected onto
inner annuli.  Spectra were extracted for 7 annuli, centered on the
cluster center and excluding the bright ridge (a sector in the range
$\rm PA= 50\degr$ -- $75\degr$).  The size of each annulus was
adjusted to obtain about 3000 photons in each ACIS-I spectrum (0.5 --
7.0\,keV).  The outer edge of the outermost annulus is about 700\,kpc
from the cluster center.  The fraction of the cluster emission in the
outermost annulus arising from gas at larger radii is corrected using
a beta model with $\beta=0.75$.  The spectral model for each annulus
is a sum of MEKAL models, all absorbed by the Galactic foreground
column density.  The parameters (temperature, abundance and norm) for
all but one MEKAL component are determined by fits to the surrounding
shells.  Parameters for the remaining component are adjusted to fit
the spectrum for the current annulus.  Deprojection results are
presented in Section~\ref{sec:sb_clump}.

\begin{figure*}[th]
\epsscale{1.1} \plottwo{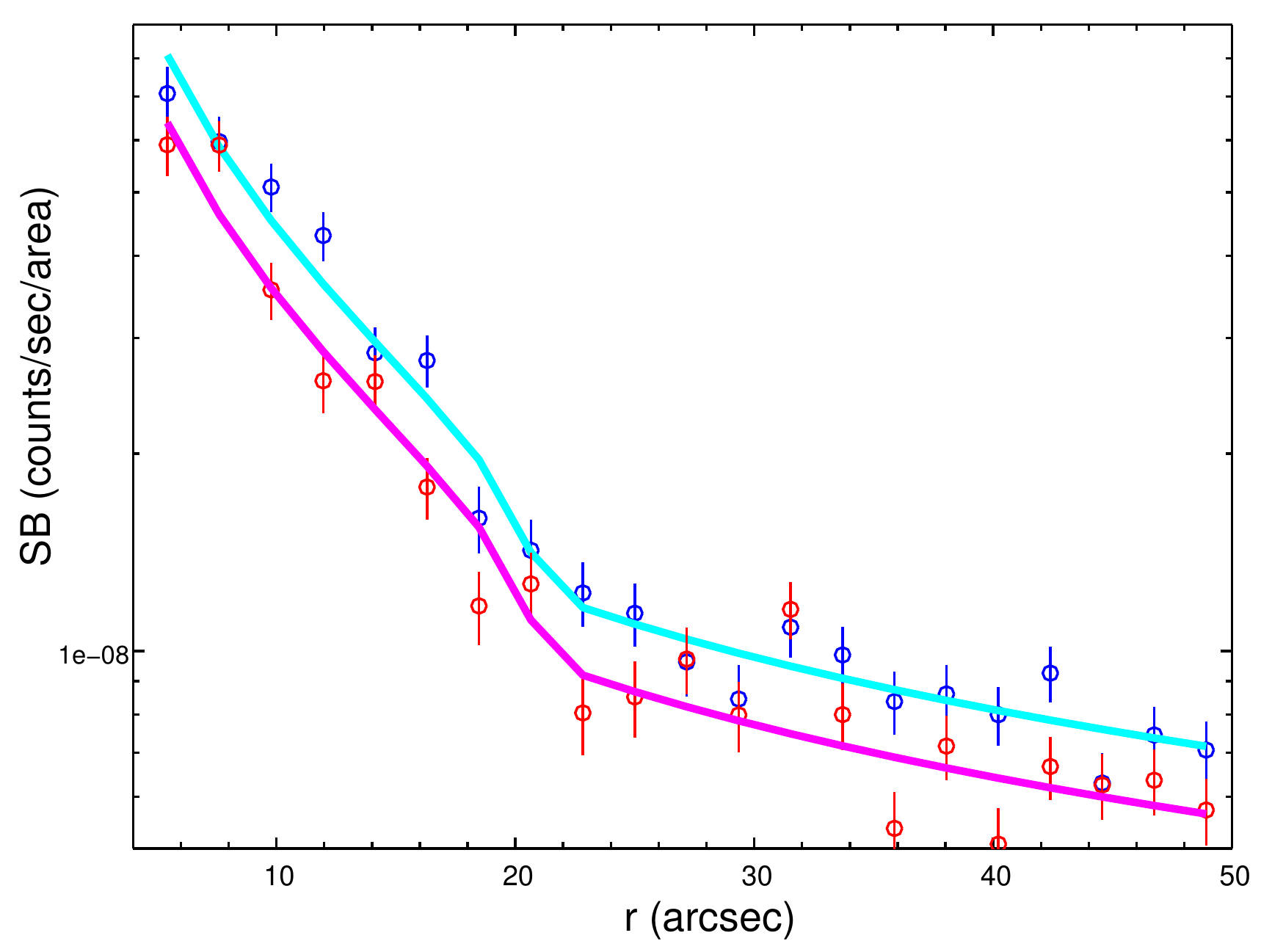}{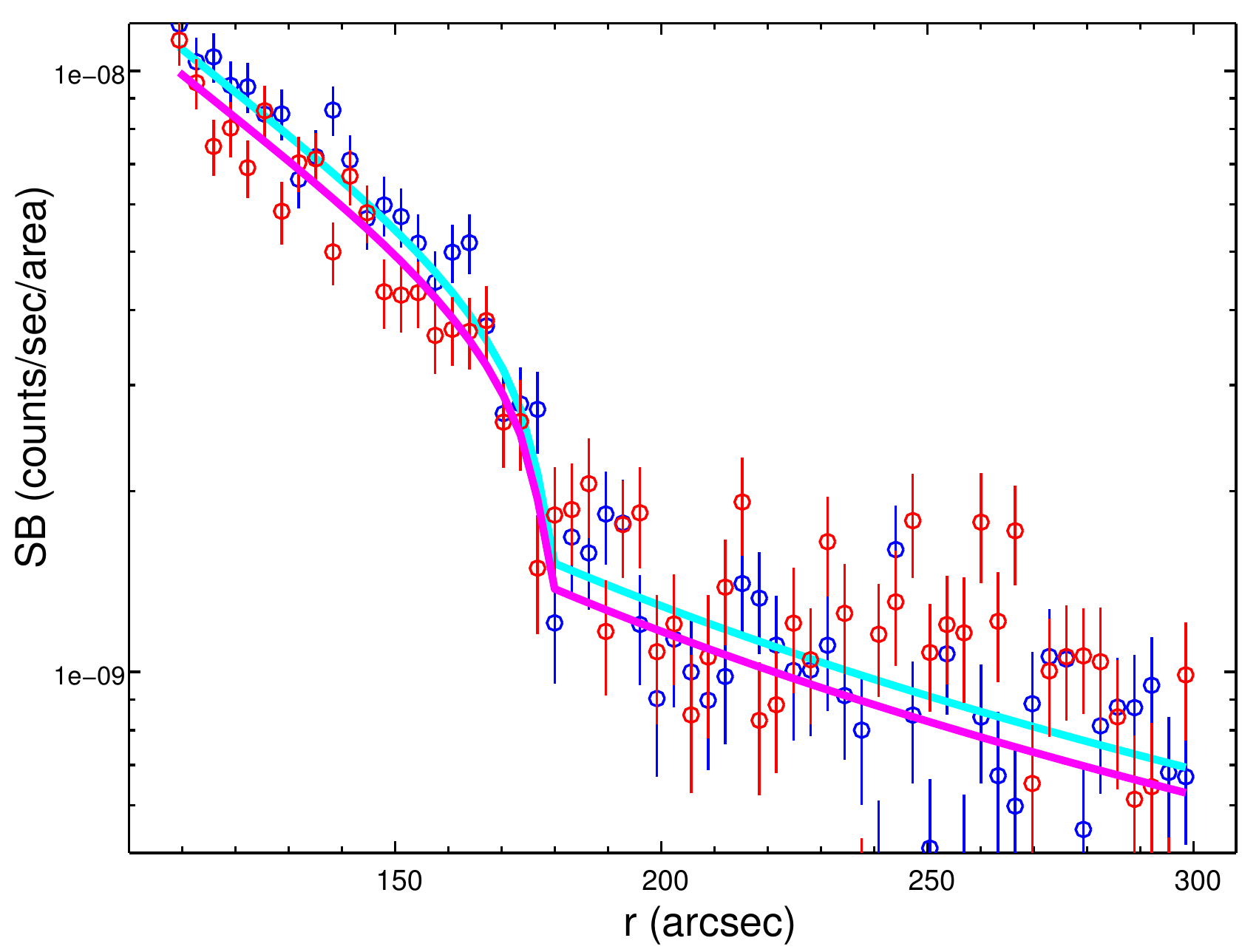}
\figcaption{Surface brightness profiles for the inner (\textit{left}
  panel) and outer (\textit{right} panel) cold fronts. The red symbols
  and magenta lines shows data points and broken power law fits for
  the ACIS-S data.  The blue symbols and cyan lines are for the ACIS-I
  data. \label{fig:sb_fronts}}
\end{figure*}

\section{Merger related structures}\label{sec:sb}

\subsection{Cold Fronts}\label{sec:sb_CF}

The new data were used to measure the surface brightness profiles and
temperature jumps for the two cold fronts to confirm their properties
with deeper data, following the procedure of \citetalias{owers09}.
Surface brightness profiles were constructed from the
point-source-free ACIS 0.5 -- 7\,keV event table, corrected for
vignetting, quantum efficiency and exposure time.  Background was
determined from scaled blank sky data.  For the northwest front, the
profile was extracted in annuli centered on $(\rm R.A., DEC) =
(168.2268\degr, 13.4344\degr)$, for PA in the range $7\degr$ --
$122\degr$.  Annuli for the southeastern front were centered at $(\rm
R.A., DEC) = (168.2197\degr, 13.4509\degr)$ with PA from $240\degr$ --
$267\degr$.  The surface brightness profile is modeled, assuming that
the electron density follows the broken power law
\begin{equation}
n_{\rm e}(r)=\left\{
\begin{array}{ll}
  n_{\rm e,1}\left({{r}\over{R_{\rm f}}}\right)^{-\alpha_1}, &  r<R_{\rm f},\\
  n_{\rm e,2}\left({{r}\over{R_{\rm f}}}\right)^{-\alpha_2}, &  r>R_{\rm f},
\end{array}
\right.
\label{eqn:density}
\end{equation}
where $R_{\rm f}$ is the radius of the discontinuity and $n_{\rm e,1}$ and
$n_{\rm e,2}$ are the inner and outer densities at $R_f$, respectively.
Here, $r$ is the spherical radius.  The small temperature dependence
of the surface brightness is ignored.  Surface
brightness  can then be determined by integrating $n_{\rm e}(r)^2$
along lines of sight.  Model parameters were obtained by jointly
fitting the ACIS-I and ACIS-S surface brightness profiles.
The density jumps were found to be $n_{\rm e,1}/n_{\rm e,2} =
1.74^{+0.26}_{-0.30}$ for the northwest discontinuity and
$2.2^{+0.27}_{-0.29}$ for the southeast discontinuity\footnote{All
  uncertainties in this section are 90\% confidence ranges.}.   

Temperature jumps were determined by jointly fitting ACIS-I, ACIS-S
and MOS spectra, using the method of \citetalias{owers09}.  Spectra
for one region on each side of a front were extracted in the sectors
defined above.  For the northwest front, the region inside the front
covered the range of radius $7\arcsec$ -- $17\arcsec$ and the region
outside the front, $23.5\arcsec$ -- $50\arcsec$.  For the southeast
front, the radial ranges are $108\arcsec$ -- $175\arcsec$ (inside) and
$190\arcsec$ -- $280\arcsec$ (outside).  The norm and temperature
outside each front were fitted using an absorbed MEKAL model, with the
metallicity fixed at the global value ($0.29$, see
Table~\ref{table:global_temp}).  Using these parameters and the
results from fitting the broken power law model for the gas density,
the norm for the gas outside the front projected onto the region
inside the front was then calculated.  This gives a two component
model for the region inside the front, one component for the gas
outside the front, with all parameters fixed, and a second component
with free parameters for the gas inside the front.  Fitting the free
parameters to the spectrum for the region inside the front then
determines the temperature of the gas there.  The resulting
temperatures are $3.15^{+0.42}_{-0.38}$\,keV, inside, and
$6.21^{+1.5}_{-0.69}$\,keV, outside the northwest front.  For the
southeast front, they are $3.29^{+0.34}_{-0.26}$, inside, and
$5.1^{+1.7}_{-0.9}$\,keV, outside.  Combining the density jumps
determined from the surface brightness profile with the temperature
jumps determined here, we find that the pressure jumps at the fronts
are $0.86^{+0.23}_{-0.22}$ for the northwest front and
$1.41^{+0.50}_{-0.48}$ for the southeast front.  These do not differ
significantly from 1, consistent with the results of
\citetalias{owers09}.  Thus, we confirm their conclusion that both
fronts are cold fronts rather than shocks. 

\begin{figure}[h]
\epsscale{1.0} \plotone{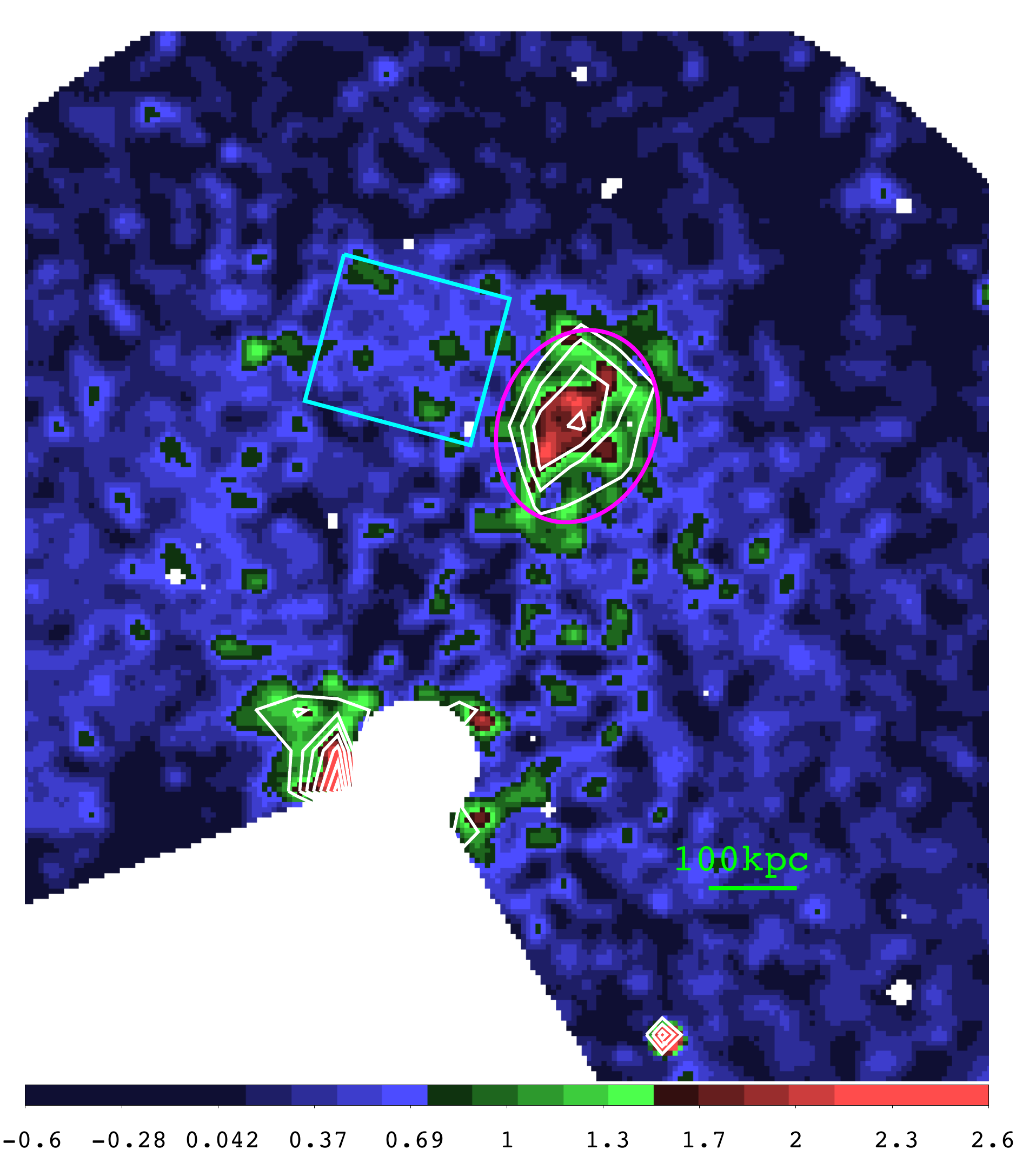}
\figcaption{Residual image after subtraction of an elliptical beta
  model from the ACIS data.  The image has been smoothed with a
  Gaussian kernel of radius $2\arcsec$. The magenta ellipse marks
  the offset core and the cyan square marks the region used to
  estimate the gas temperature and density of the
  tail.\label{fig:densitypfl_resid}}
\end{figure}

\begin{figure*}[t]
\epsscale{1.0}
\hspace{3mm}\includegraphics[scale=0.95,bb = 0 0 500 360]{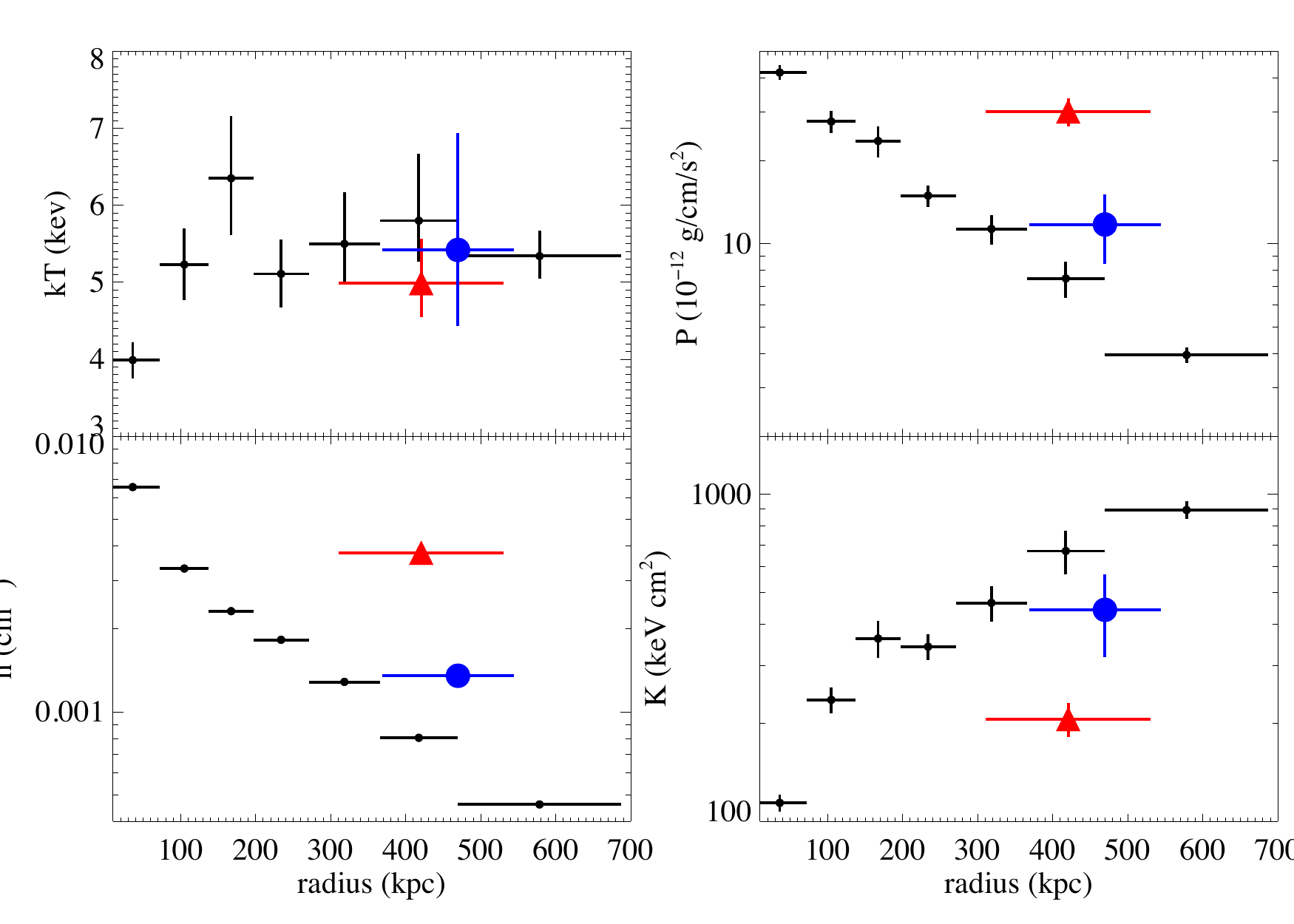}
\caption{Deprojected temperature, density, pressure, and entropy
  profiles.  Values are shown for the offset core (red triangle) and the tail
  (blue sphere) at the radii of the regions outlined in
  Figure~\ref{fig:densitypfl_resid}.  The uncertainties for some
  points are too small to be seen. \label{fig:densitypfl}}
\end{figure*}

\subsection{Offset Core}\label{sec:sb_clump} 

Using a  KMM \citep[KayeÕs mixture model;][]{ashman94} analysis, 
\citetalias{owers09} divided the 3-dimensional projected phase space of cluster members into a sum of Gaussians. Under this partitioning, \citetalias{owers09} found a compact substructure at the location of
the offset gas core, which is marked by the ellipse in
Figure~\ref{fig:imagexmm}.  This substructure has a velocity of
432\,km\,s$^{-1}$ relative to the cluster mean and a velocity
dispersion of 166\,km\,s$^{-1}$.  Although the X-ray emission from
this region is not clearly distinct from the core of A1201, a
pronounced ridge of X-ray emission extends from the core to the
position of the substructure.  We interpret this ridge as being
composed of contributions from the large-scale cluster emission of
A1201 and from gas associated with the offset core.  In order to
highlight emission from the offset core, a residual image, made by
subtracting an elliptical beta
model\footnote{http://cxc.harvard.edu/sherpa4.2.2/ahelp/beta2d.html}
fitted to the large-scale emission from the combined ACIS image, is
shown in Figure~\ref{fig:densitypfl_resid}.  The cool core and the
sector containing the southeast cold front were both masked out of the
fit, since they are also not well-described by the beta model.  The
offset core stands out clearly at the end of the ridge.

The temperature and density of the offset core were estimated using
spectra for the region marked by the magenta ellipse in
Figures~\ref{fig:imagexmm} and \ref{fig:densitypfl_resid}.  A local
background was extracted from an annulus centered at the cluster
center and containing the offset core, in the PA range $0\degr$ --
$50\degr$.  Scaling this to the area of the ellipse to obtain a
background spectrum and fitting an absorbed thermal model to the
residual spectrum gives the excess emission measure of the offset core
as $1.145^{+0.085}_{-0.077}~\times 10^{66}~{\rm cm^{-3}}$.
For a fixed emission measure in a fixed volume, the mean gas density
is maximized when it is uniform.  Assuming that the ellipse is the
projection of a spheroid, with semi-major axis $a=110$\,kpc,
semi-minor axis $b=88$\,kpc, and its third semi-axis $c =
\sqrt{ab}$, the electron density of the gas is ${n} =3.4\pm0.2\times
10^{-3}\,{\rm cm}^{-3}$.  The uncertainty here is dominated by
systematic errors, particularly in the volume of the emitting region.
The errors show the range of densities as the third axis, $c$, is
varied from the prolate ($c=b$) to the oblate ($c=a$) extremes.  The
corresponding gas mass of the offset core is $3.8\pm0.2\times
10^{11}\,M_{\sun}$.  The bright galaxy at $\rm (R.A., DEC) =
(168.2088, 13.4750)$ is dynamically associated with the offset core,
located close to its center, and has a stellar mass, estimated from
SDSS photometry, of ${\rm M_{gal}} = 3\times 10^{11} M_{\sun}$, which
is comparable to the gas mass.  The second brightest galaxy associated
with the offset core is one magnitude fainter.  Only taking into
account the stellar mass of the brightest galaxy and assuming that the
ratio of the total mass to stellar mass is about 10 gives a
conservative lower limit on the total mass of the offset core of
$3\times10^{12} M_{\sun}$. 

The temperature maps show that the offset core is only marginally
cooler than surrounding regions (Figure~\ref{fig:matt_hires_tmap}).  Since
the offset core is brighter than adjacent regions at the same radius,
it must be denser, as confirmed by the results above.  In
Figure~\ref{fig:densitypfl}, results for the temperature, density,
pressure and entropy of the gas in the offset core are plotted in red
together with the results of the deprojections. We note that the
bright ridge was excluded from the spectra used for the deprojections,
although this makes very little difference to the results. We also obtained a deprojected metallicity profile, which is consistent with the metallicity map in the right panel of Figure~\ref{fig:matt_tmap}. The metallicity is highest in the cool core, $Z=0.86\pm0.2$ within 72\,kpc of the cluster center, and drops to $Z\sim0.3-0.4$ at 140 to 690\,kpc. 

Although the gas temperature of the offset core is only marginally
lower than that of the surrounding gas (top left panel of
Figure~\ref{fig:densitypfl}), its entropy is significantly lower.
This shows that the high density of the gas within the offset core is
not simply the result of adiabatic compression of surrounding gas,
since that would not alter the entropy.  Thus, the gas in the offset
core must originate in some other place.  The simplest interpretation is
that it is gas that fell in with the core. The abundance map in the right panel of Figure~\ref{fig:matt_tmap} shows marginal evidence of lower abundances near the offset core. While the significance of this feature is low, it is consistent with an origin for this gas outside the present cluster.

We can use the pressures to make a rough estimate of the direction of
motion of the offset core.  The thermal pressure of gas in the offset
core is balanced by the external thermal pressure plus the ram
pressure at the stagnation point.  Using our estimates of the density
and temperature from above gives an internal pressure for the core of
$p_{\rm core} = 3\times10^{-11}\,{\rm dyne\,cm^{-2}}$.  The
pressure of the external gas at the same radius is $p_{\rm ext}$ is
$10^{-11}\,{\rm dyne\,cm^{-2}}$. Therefore, the ram pressure $p_{\rm rp}$ is about
$2\times10^{-11}\,{\rm dyne\,cm^{-2}}$.  Since the ram pressure
$p_{\rm rp} = \rho_{\rm ext} v^{2}$, the velocity, $v$, of the offset
core would need to be about $1000$\,km\,s$^{-1}$.  

Pressure arguments also constrain the location of the offset core
along our line of sight.  Both the thermal pressure and the ram
pressure of the external gas depend on its density, which decreases
with distance from the cluster center.  If the distance of the offset
core from the cluster center exceeds the projected separation, then
the external gas pressure would be smaller, increasing the ram
pressure required to confine the core.  Since the gas density is also
lower, a larger speed is then demanded to confine the gas in the
offset core.  However, in the absence of any evidence for a shock
\citepalias{owers09}, we assume that the motion of the core is subsonic,
i.e., its speed cannot exceed $\sim1100\rm\,km\,s^{-1}$, which is not
significantly higher than the velocity estimated above.  Thus, the real
distance of the offset core from the center of A1201 cannot be much
greater than the projected separation of $\simeq 420$\,kpc.  This
requires the offset core to be located close to the plane of the
sky.

\begin{figure*}[t]
\epsscale{0.5} \plotone{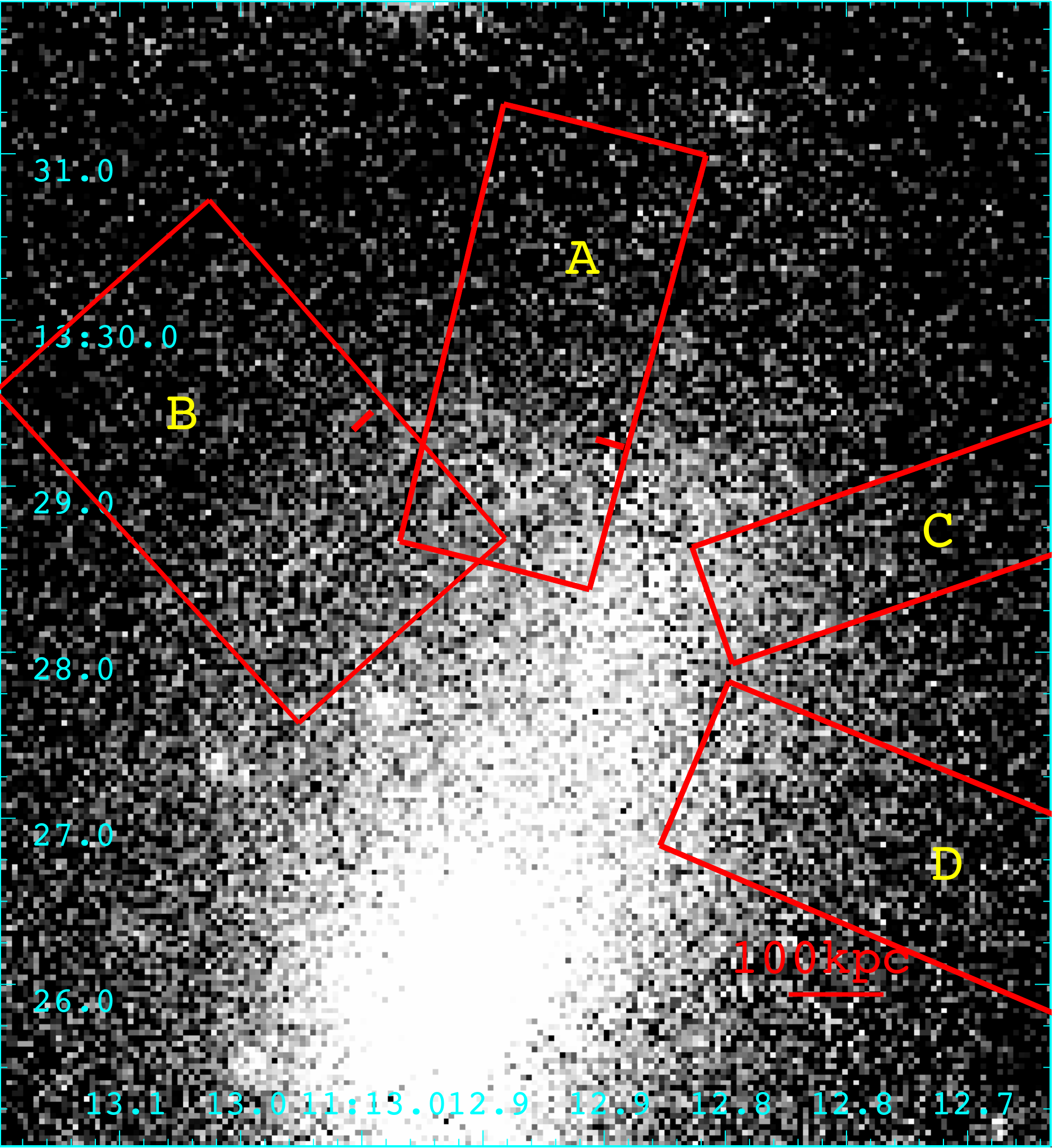}
\epsscale{0.6} \plotone{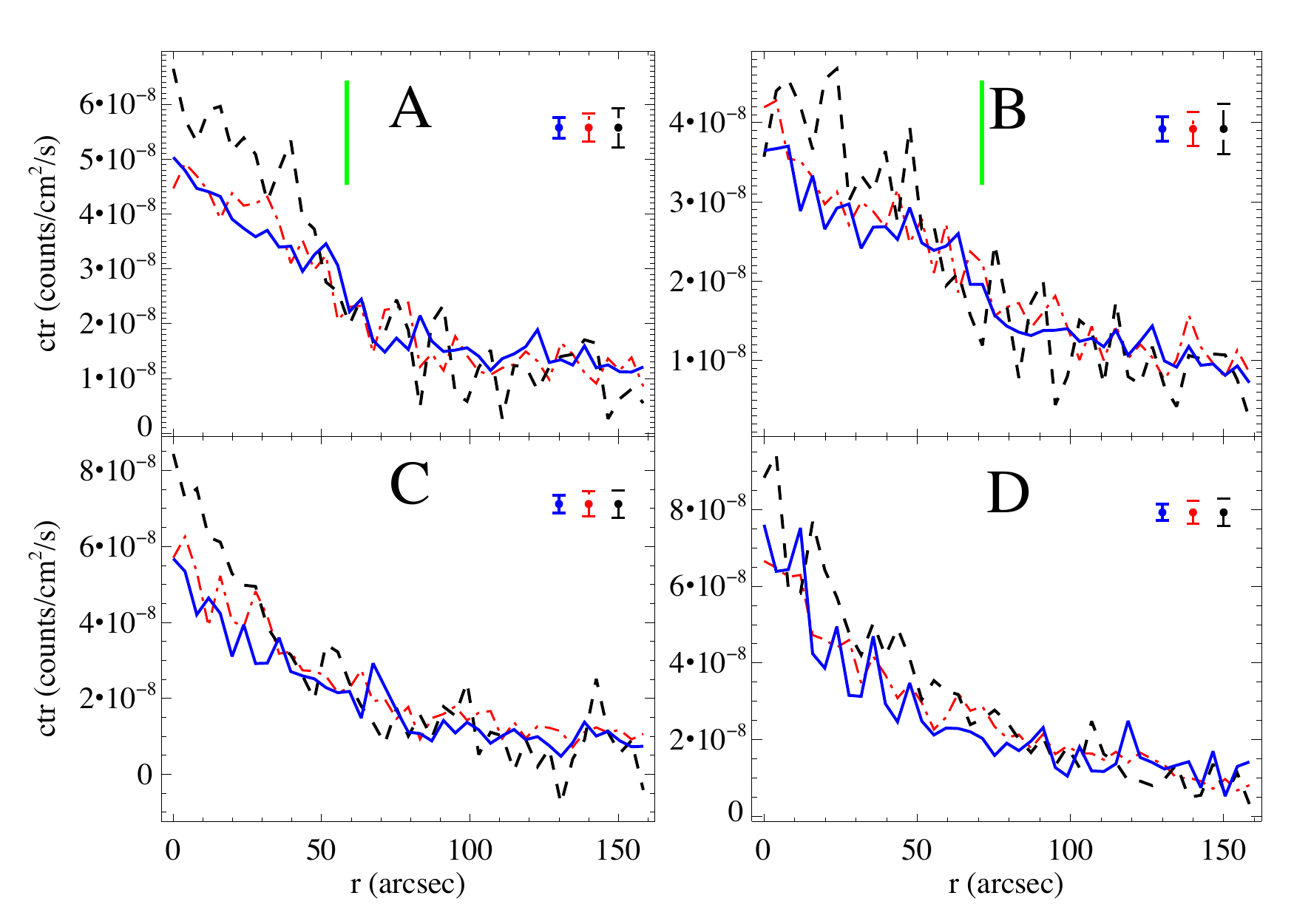}
\figcaption{{\em Top:} EPIC image zoomed in on the northern tail.  The
  four boxes show the regions used to extract surface brightness
  profiles shown in the lower panels: A, northern edge; B, eastern
  edge; C and D comparison regions.  The locations of the edges are
  marked with ticks in regions A and B. {\em Bottom:} Surface
  brightness profiles for the four regions.  Radius increases away
  from the cluster center in each panel, from zero at the inner edge
  of each box.  Profiles for the merged ACIS data are plotted with
  black dashed lines, for the PN data with blue solid line, and for
  the MOS data with red dashed-dotted lines.  The locations of the
  edges marked in green in the lower panels for boxes A and B
  correspond to the ticks in the boxes of the top panel.  Typical
  statistical uncertainties are shown at the top right of each
  panel.  \label{fig:sb_tail}}
\end{figure*}

\subsection{Tail of the Offset Core}\label{sec:sb_tail}

In the two panels of Figure~\ref{fig:imagexmm}, there is a region of
brighter emission running to the northeast of the offset core that has
a relatively sharp boundary to the north and a less well-defined edge
to the east (marked by the green curve).  We interpret the bright
emission as a stream of gas being stripped from the offset core.
Surface brightness profiles, obtained by the procedure of
Section~\ref{sec:sb_CF}, for the northern edge of this feature (from
box A in the upper panel of Figure~\ref{fig:sb_tail}) and its eastern
edge (from box B) are shown in the lower panels of
Figure~\ref{fig:sb_tail}.  The surface brightness profile for box A
shows a significant drop across the edge in all three data sets.  This
feature was not seen by \citetalias{owers09} because of its limited
coverage in the ACIS-S data, where the tail lies close to a chip gap.
An abrupt drop in surface brightness can also be seen in the PN data
(blue solid lines in the bottom panels of Figure~\ref{fig:sb_tail}) at
the eastern edge of this feature, in the surface brightness profile
for box B.  The eastern edge is less significant in the noisier MOS
and ACIS data, although they are reasonably consistent with the PN
profile. Without the ACIS identification of the edge, the exact
location of the eastern edge is not as securely constrained as the
northern one.  Profiles extracted from boxes C and D, on the opposite
side of the bright ridge are shown for comparison.  In them, the
surface brightness fades smoothly away from offset core.  Excess
emission from the tail can also be seen in the residual map of
Figure~\ref{fig:densitypfl_resid}.

In addition to the excess X-ray emission, the distribution of the
galaxies associated with the offset core by the KMM algorithm is
elongated in the direction of this tail (top panel of Figure 14 in
\citetalias{owers09}).  This is suggestive of a tidal tail of galaxies
stripped from the disrupting core.  However, the small number of
galaxies in this structure, combined with the uncertainties in the KMM
decomposition, make this hard to demonstrate with any certainty.

Although the tail does not stand out in the temperature or abundance maps of
Figure~\ref{fig:matt_hires_tmap} and \ref{fig:matt_tmap}, both the temperatures and abundances are lower on the eastern side of the offset core, over the tail than they are to west. The absence of a significantly lower temperature in the tail may be simply because the stripped gas
in the tail does not dominate the X-ray emission.  To examine the
properties of the tail more carefully, a spectrum was extracted from
the tail region, defined as the rectangular region marked in
Figure~\ref{fig:densitypfl_resid}.  Apart from scaling, the local
background used is the same as that used to measure the spectrum of
the offset core in Section~\ref{sec:sb_clump}, i.e., a sector from PA
$0\degr$ to $50\degr$ of the annulus containing the offset core.
X-ray emission in this region is dominated by the ICM of A1201, so
that, after background subtraction, the spectrum should be dominated
by emission from the tail.  These spectra for the ACIS and MOS data
were fitted jointly using the absorbed thermal model of
Section~\ref{sec:data_spec}.  The temperature of the tail was found to be
$5.42^{+1.5}_{-1.0}$\,keV.  To estimate the density of the tail, we
assume that the rectangular region from which the spectrum was
extracted is the projection of a cylinder, with its height defined by
the long side of the rectangle and its diameter by the short side.
This gives an electron density for the tail of
$1.31\pm0.1\times10^{-3}$\,cm$^{-3}$, where only the
statistical uncertainty (90\% confidence level) is included.  Although
the systematic error is large, the density of gas in the tail is
significantly lower than in the offset core (consistent with its lower
surface brightness), but 1.7 times greater than the ICM density at the
same radius (Figure~\ref{fig:densitypfl}).

The temperature and density determined here imply that the gas
pressure of the tail is appreciably greater than that of its
surroundings.  In that case, the tail should expand at close to its
internal sound speed, until it comes into pressure equilibrium with
the surrounding gas.  For the sound speed at $\sim 5$\,keV and a tail
width of $\simeq 100$ kpc, this only requires $\simeq 0.1$\,Gyr.
Taking the velocity of the offset core to be about $700$\,km/s (see
next section), the core would travel about 70\,kpc in this time, which
should determine the length of the tail (projection effects can only
reduce this).  However, the extent of the tail is considerably more
than 70\,kpc (see Figure~\ref{fig:sb_tail}).  The most plausible cause
of this discrepancy is that our estimates of the temperature and
density of the tail do not accurately reflect its mean temperature and
density.  At typical cluster temperatures, X-ray brightness in
detectors like ACIS and EPIC is quite insensitive to the gas
temperature, being largely determined by its emission measure.  Thus,
our density estimate best represents the root-mean-square (RMS)
density of the tail.  Note that the excess brightness of the tail
demands that its RMS density is higher than that of the surroundings.
Stripped gas in the tail could be clumpy, making the RMS density
larger than its mean density.  Coupled to this, the temperature in the
gas would almost certainly be non-uniform as well.  Temperatures
obtained by fitting single thermal models to emission from gas
mixtures are not simply related to the mean temperature.  Thus, our
estimates of both of the temperature and density in the tail may not
provide a good estimate of its mean pressure.  It is more physically
reasonable that most of the tail is close to local pressure
equilibrium.  Presumably, the gas is in the process of mixing into the
ambient ICM.  We note that to make the RMS density 1.7 times higher
than the mean density, the distribution of gas densities in the tail
would need to be broad.  We also note that pressure estimate for
the offset core may be affected similarly (in which case, the velocity
estimate of section \ref{sec:sb_clump} would be high).  However, because it
stands out more clearly relative to the local cluster X-ray emission,
our pressure estimate for the offset core should be less affected.

\section{Discussion}\label{sec:discussion} 

\begin{figure}
\plotone{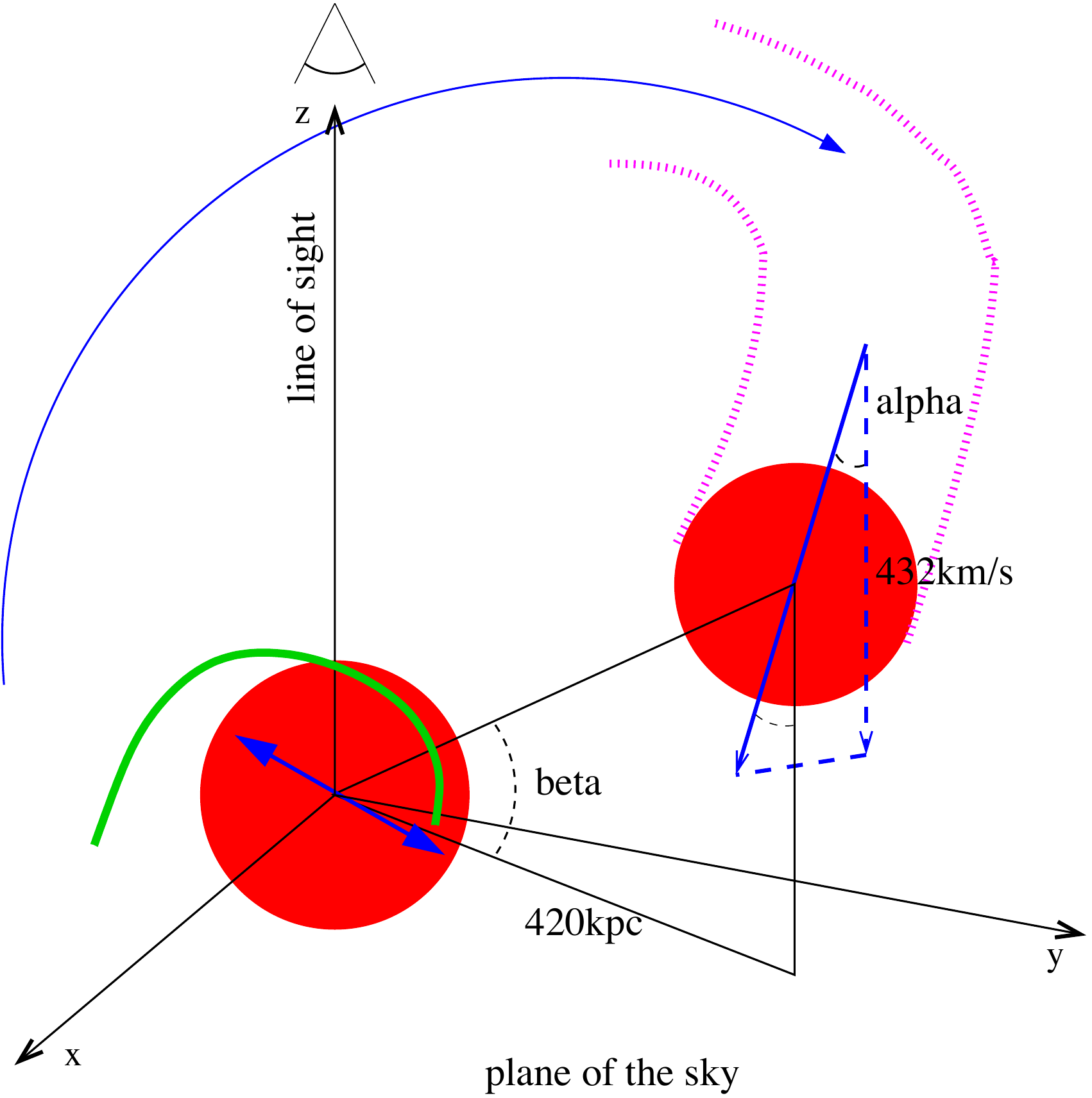} 
\figcaption{A simple merger model for A1201.  Our line of sight is
  along the $z$ axis, from the $+z$ direction, while the $x-y$ plane
  is the plane of the sky.  The sphere at the origin represents the
  core of the primary cluster and the other sphere represents the merging
  core.  The blue solid vector from the offset core indicates
  its current direction of motion.  The offset core is trailed by
  stripped gas within magenta dashed curves.  The green curve represents the spiral front created by
  the sloshing motion excited in the merger. The blue arrows indicate the motion of the two cores. \label{fig:scheme}}
\end{figure}

We begin by presenting our interpretation of the merger history of
A1201 (Figure~\ref{fig:scheme}).  On its initial core passage, the
infalling subcluster was approaching us along a path to the southeast
of the center of A1201 in projection.  Much of the gas belonging to
the subcluster would have been stripped during its first pericenter
passage, although some gas has been retained by the subcluster core.
The subcluster excited sloshing motions in the core of the primary
cluster during its first core passage \citep{markevitch07}, giving
rise to the cold fronts.  The remnant subcluster core has now passed
apocenter and it is close to its second pericenter passage, moving
away from us.  The orbit of the subcluster is roughly in a plane
containing our line of sight ($-z$ direction in the figure) and the
X-ray bright ridge.
The evidence supporting this interpretation is outlined below.

\begin{figure}
\plotone{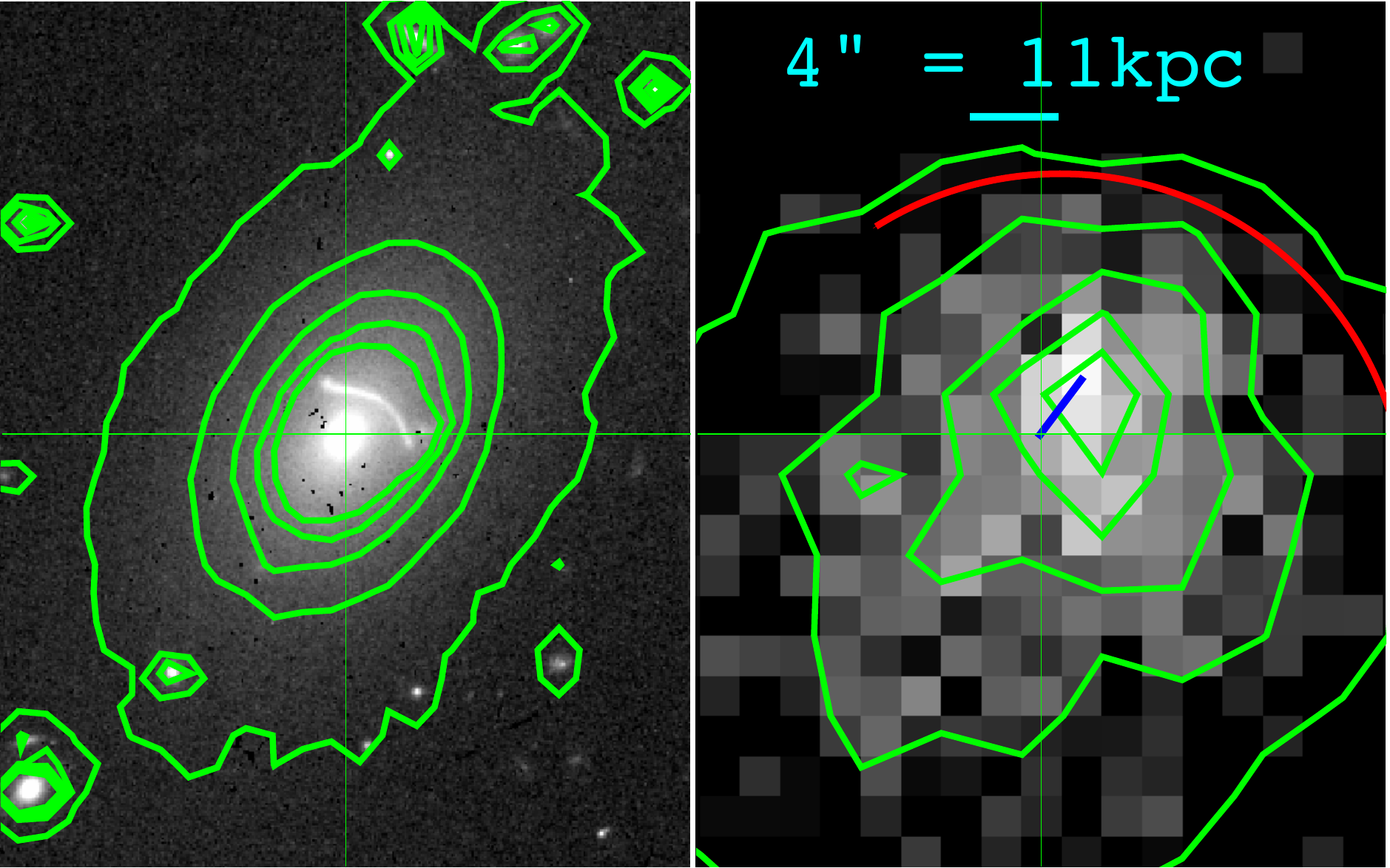} 

\figcaption{WFPC2 image of the BCG (left) and X-ray image of the
  cluster core (right).  The WFPC2 image was made with the F606W
  filter 
\citep{edge03}.  Surface brightness contours shown in each panel apply
to the image in the panel.  The red curve in the right panel marks the
northwest cold front discussed in Secton~\ref{sec:sb_CF}.  The cross
marks the center of the BCG, which is offset from the peak of the
X-ray emission by about 11\,kpc in projection.  The major axis of the
optical image is aligned with the direction of offset of the X-ray
peak and the major axis of the X-ray emission. \label{fig:hst_bcg}}
\end{figure}

\renewcommand{\labelitemi}{$\bullet$}

\subsection{Orientation of the Orbit}

  Simulations of sloshing cold fronts show that contact
  discontinuities form along a spiral of enhanced X-ray emission that
  grows in the plane of the interloper orbit \citep{ascasibar06,
    poole06}.  This feature is generally visible in X-ray images,
  unless the merger is viewed from close to the plane containing the
  spiral.  Viewed from within its plane, the spiral appears as cold
  fronts on alternating sides of the cluster center.  Since no spiral
  feature can be seen connecting the fronts in A1201, in either the
  X-ray image (Figure~\ref{fig:imagexmm}) or the temperature maps
  (Figure~\ref{fig:matt_hires_tmap}), it seems likely that we are viewing
  the spiral front from close to its plane, i.e. from close to the
  plane of the merging orbit.

  In the X-ray images of A1201, a bright ridge lies along the axis
  joining the two sub-cluster cores.  At least in part, this is
  composed of emission from the sloshing gas and the offset core.  The
  two cold fronts are projections of the spiral feature, viewed from
  roughly within the plane of the spiral.  In this orientation,
  enhanced X-ray emission from the whole of the spiral density feature
  (not just the cold fronts) is projected onto the nearly edge on
  plane of the orbit, contributing to the X-ray emission from the
  bright ridge.

 The simulations show that the spiral fronts created by sloshing
  wind from the outside inward in the prograde sense of the interloper
  orbit \citep[e.g.][]{ascasibar06,poole06}.  If the direction of the
  tail reveals the earlier path of the offset core, we expect that the
  core passed through its pericenter on the southeastern side of the
  center of the primary cluster, passing around it moving towards the northeast in
  projection.  Thus, the spiral structure should wind from the inside
  out, counterclockwise, as plotted in Figure~\ref{fig:scheme} and as
  projected onto the sky.  
The X-ray structure between the cold fronts is consistent with this scenario (Figure~\ref{fig:matt_tmap}).  
  We note that in simulations the
  outermost front generally appears on the opposite side of the
  cluster center to the pericenter passage that excited the sloshing
  \citep[e.g.][]{roediger11}, whereas the outer cold front lies to the
  southeast in A1201.  This might be an issue for our proposed merger
  orbit.  However, in the simulation shown in Figure~7 of
  \citet{ascasibar06}, the outer cold front changes sides at later
  times.  This may well be a consequence of the second core passage.
  For our purpose, it only matters that these simulations demonstrate
  that the location of the outer front does not unambiguously
  determine the side of first core passage.  This issue is further
  complicated by projection.

\subsection{Age of the Merger}

  We can estimate the time elapsed since pericenter passage using
  the locations of the two cold fronts.  First, if the sloshing motions
  that give rise to cold fronts may be regarded as a superposition of
  internal gravity waves excited by the merging subcluster
  \citep{churazov03}, we can estimate the time required for the cold
  fronts to get out of phase by $\pi$ radians as $\tau =
  \pi/(\omega_{\rm BV, in} - \omega_{\rm BV, out})$, where
  $\omega_{\rm BV}$ is the Brunt-V\"ais\"al\"a frequency
  \citep{owers11b, simionescu11}.  This is given
  by 
\begin{equation} \label{omega_BV}
\omega_{\rm BV}=\sqrt{{g \over r} {3\over 5} {d\ln \Sigma \over d\ln r}} = 
\omega_{\rm K} \sqrt{{3\over 5} {d\ln \Sigma \over d\ln r}},
\end{equation}
where $r$ is the radius, $g = GM(r)/r^2$ is the acceleration due to gravity, $\Sigma$ is the entropy index and $\omega_{\rm K} =
\sqrt{g/r}$ is the Kepler frequency.  From observations
\citep[e.g][]{owers11b}, the factor under the square root on the right
is close to unity, so that the approximation $\omega_{\rm BV} \simeq
\omega_{\rm K} = V_{\rm circ}/r$ is adequate for our purposes ($V_{\rm
  circ} = \sqrt{gr}$).  To estimate $\omega_{\rm K}$, we measure radii
from the cluster center, which is assumed to mark the potential
minimum, and we assume that the mass distribution can be approximated
as a singular isothermal sphere, so that the circular velocity is
given by $V_{\rm circ} = \sqrt{2}\sigma_{v}$, where $\sigma_v$ is the
line of sight velocity dispersion.  In A1201, the radii of the two
cold fronts are 300\,kpc and 50\,kpc, and $V_{\rm circ} =
1100$\,km\,s$^{-1}$, giving $\tau \simeq
0.17\rm\,Gyr$. 
The Brunt-V\"ais\"al\"a frequency applies to purely transverse modes. For wavevectors with a radial component, the frequency is reduced by a factor of the sine of the inclination to the radial direction. Cold fronts are oriented almost radially, so this correction can be substantial.  
\citet{owers11b} and \citet{roediger11} found that the approximation 
in Eqn.~\ref{omega_BV} underestimates the time since pericenter passage by a factor of
3 -- 4.  Correcting for this, we find that the first pericenter
passage occurred $\sim 0.6$\,Gyr ago.  This age is smaller than expected
if the merging core is close to its second pericenter passage.  Note,
however, that the 6:1 ratio of the radii of the two cold fronts in
A1201 is substantially larger than the 4:1 ratio for the Virgo cluster
fronts modeled by \citet{roediger11}.  Since the estimate above is
largely determined by the location of the inner front, it may well be low.

The location of the outer front provides an alternative estimate of
time elapsed since pericenter passage, assuming that the outer front
was excited then.  \citet{roediger11} found that cold fronts propagate
outward with constant speed.  For their Virgo cluster model, the outer
front takes 1.5\,Gyr to reach a radius of $\sim90$\,kpc.  Given that
A1201 is about twice as hot as the Virgo cluster, its characteristic
speeds (sound speed and Kepler speed) should be a factor of $\simeq
\sqrt{2}$ greater than for Virgo.  The speed of the cold fronts scales
with these, so that, scaling from the results of \citet{roediger11},
the outer front in A1201 would require about 3.5\,Gyr to reach its
current radius of 300\,kpc.  This estimate leaves comfortably enough
time for the merging subcluster to have reached its second pericenter
passage.  A more accurate estimate for the age of the merger requires
a simulation better matched to A1201. 

These arguments show that sufficient time has elapsed
for the merging subcluster to have returned for its second core
passage.

\subsection{Dynamics of the Offset Core}

  As discussed in section \ref{sec:sb_clump}, we estimate the
  speed of the offset core to be about $1000$\,km\,s$^{-1}$, while the
  line of sight velocity of the core is 432\,km\,s$^{-1}$.  To be
  consistent, these values require the core to be moving at an angle
  of about $65\degr$ to our line of sight ($\alpha$ in
  Figure~\ref{fig:scheme}).  We note that this angle is distinct from the
  angle between the orbital plane of the offset core and the line of
  sight, as the direction of motion of the core within the orbital
  plane is not constrained (Figure~\ref{fig:scheme}).  A relatively
  large value of $\alpha$ thus presents no conflict with the proposed
  model, which requires our line of sight to be nearly parallel to the
  orbital plane.  The argument of \S\ref{sec:sb_clump} assumes that
  the core is now subsonic.  Note that if it was moving much faster
  than $1000$\,km\,s$^{-1}$, its low line of sight velocity would
  require its motion to be even closer to the plane of the sky,
  increasing the likelihood that we could see the shock
  front it would create.  This favors our assumption that its speed
  through the cluster is subsonic or, at least, only mildly
  supersonic.

  As also argued in section \ref{sec:sb_clump} based on pressures,
  the distance of the offset core from the center of A1201 cannot be
  much greater than the projected separation of $\simeq 420$\,kpc.
  This requires the offset core to be located reasonably close to the
  plane of the sky.  In other words, the angle $\beta$ in
  Figure~\ref{fig:scheme} cannot be large.

   The relatively high pressure in the offset core
requires it to lie fairly close to the plane of the sky.  Thus, we
interpret the offset core as the remnant of a merging subcluster,
close to its second core passage.

\subsection{Alignment of the Sub-Structures}

\citet{edge03} used the strong lensing arc projected onto the BCG
(Figure~\ref{fig:hst_bcg}) to calculate the mass distribution in the
cluster center.  They found the mass distribution to be elongated
along the same axis as the BCG isophotes, but, interestingly, with a
significantly greater ellipticity than the isophotes.  Unfortunately,
they found no lensed arcs further from the cluster center, nor enough
galaxies for a weak-lensing analysis.  Thus, their mass model is
well-constrained only in the cluster core and the extent of the highly
elliptical mass distribution is not known.  At their time of
publication, no high resolution X-ray image of A1201 had been
obtained, leaving them unaware of the substructures discussed
here.  

The X-ray data combined with the optical data of \citetalias{owers09}
make a strong case for the presence of a remnant core at a distance of
about $400$\,kpc from the cluster center, in the direction of
elongation of the mass distribution.  This core is too far from the
cluster center to contribute significantly to the mass distribution
there.  However, the disturbance in the dark matter and stars created
by the merging core might well have elongated the mass distribution in
this direction.  It is remarkable that the major axis of the mass
distribution, the offset between the BCG and the X-ray peak, and the
cold fronts all lie along the same direction. Conventionally, alignments between cluster halos and their satellites are controlled by the orientation of large scale structures \citep[e.g.][]{basilakos06}. However, the large scale structure should not play an important role in A1201, as the location of the interloper changes rapidly at this stage of merging. Thus, the consistency of all alignments shown in A1201 suggests the possibility that the orientation of the BCG and the mass distribution in the cluster core are affected by the satellite directly.  This would be surprising, since the mass in the cluster core should not be affected much by the tidal field of the interloper \citep[cf.][]{faltenbacher08}.  We speculate that alignment may be enhanced by sloshing, which would need to be verified by numerical simulations.

\section{Summary}\label{sec:summary}

We have analyzed the structure and dynamics of the merging cluster A1201
using the \xmm{} and \chandra\ data. Structures associated with an infalling cluster, including cold fronts and an offset remnant core, were identified previously by ONC. 
In addition to these structures, the new data show enhanced emission east of the remnant gas core, with breaks in surface brightness along its boundary to the north and east. This is interpreted as a tail of gas stripped from the offset core.

Temperature and metallicity maps of A1201 made from the new data support the merger interpretation. High metallicity in the cool core which drops abruptly across the southern cold front is consistent with a sloshing cold front. There is a hint of reduced metallicity in the offset core and tail, as expected if these arise in an external system. Using the deprojected density, temperature, and entropy profiles of the cluster, the entropy of the  gas in the offset core and the tail were found to be lower than in the gas at the same radius elsewhere in the cluster.  This evidence is consistent with the offset core being the remnant gas of a merging satellite cluster and the tail being composed of gas stripped from it.  

The observed properties of this system, including the placement of the
cold fronts, the offset core and its tail, together with our estimate
for the velocity of the offset core are consistent with a simple
merger model for A1201 that is sketched in Figure~\ref{fig:scheme}.
In this model, the offset core passed pericenter to the southeast of
the primary cluster core and it is now on its second pericenter passage, moving with a
transonic velocity of $\sim 1000\rm\,km\,s^{-1}$. The compact, marginally low temperature, structure of the offset core indicates that this gas belongs to the interloper, having survived the first core passage and the gas is being stripped to form the tail behind it. 
The gas in the primary core was perturbed by the infalling core, causing
the sloshing that gave rise to the two visible cold fronts.  The disposition of the cold fronts requires that the merger is viewed from close to the plane containing the orbit of the interloper. Moreover, the remarkable alignment between the major axis of the mass distribution
in the cluster core, the offset between the BCG and the X-ray peak,
the cold fronts and larger scale structure suggests that they have all
been affected by the merger disturbing the core of the primary
cluster. 

\acknowledgements  This work was partly supported by NASA grants
NAS8-03060 and NNX08AD68G.  CJM and BRM are supported by Chandra Large Project Grant: G09-0140X. BRM acknowledge generous support from the Natural Sciences and Engineering Research Council of Canada. MSO and WJC acknowledge the financial support of the Australian
Research Council. We have made use of data obtained under
the Chandra HRC GTO program and software provided by the Chandra X-ray
Center (CXC) in the application packages CIAO, ChIPS, and Sherpa.
STSDAS is a product of the Space Telescope Science Institute, which is
operated by AURA for NASA.

\end{document}